\documentclass[11pt,floatfix,showpacs]{revtex4}
\usepackage{amsmath,amssymb}
\usepackage{graphicx}

\def\bea{\begin{eqnarray}}
\def\eea{\end{eqnarray}}
\def\beq{\begin{equation}}
\def\eeq{\end{equation}}

\begin{document}
%\preprint{TAN-FNT-03-xx}

\title{Meson properties at finite temperature in a three flavor nonlocal chiral quark model
with Polyakov loop}

\author{G.A. Contrera$^{a,b}$, D. G\'omez Dumm$^{b,c}$ and
Norberto N. Scoccola$^{a,b,d}$}

\affiliation{
$^a$ Physics Department, Comisi\'on Nacional de Energ\'{\i}a At\'omica, \\
 Av.Libertador 8250, (1429) Buenos Aires, Argentina.\\
$^b$ CONICET,
Rivadavia 1917, 1033 Buenos Aires, Argentina\\
$^c$ IFLP, Dpto.\ de F\'{\i}sica, Universidad Nacional de La Plata,
     C.C. 67, (1900) La Plata, Argentina.\\
$^d$ Universidad Favaloro, Sol{\'\i}s 453, (1078) Buenos Aires,
Argentina.}

\begin{abstract}
We study the finite temperature behavior of light scalar and
pseudoscalar meson properties in the context of a three-flavor
nonlocal chiral quark model. The model includes mixing with active
strangeness degrees of freedom, and takes care of the effect of
gauge interactions by coupling the quarks with the Polyakov loop.
We analyze the chiral restoration and deconfinement transitions,
as well as the temperature dependence of meson masses, mixing
angles and decay constants. The critical temperature is found to
be $T_c \simeq 202$~MeV, in better agreement with lattice results
than the value recently obtained in the local SU(3) PNJL model. It
is seen that above $T_c$ pseudoscalar meson masses get increased,
becoming degenerate with the masses of their chiral partners. The
temperatures at which this matching occurs depend on the strange
quark composition of the corresponding mesons. The topological
susceptibility shows a sharp decrease after the chiral transition,
signalling the vanishing of the U(1)$_A$ anomaly for large
temperatures.
\end{abstract}


\maketitle
\section{Introduction}

The detailed understanding of the behavior of strongly interacting matter
under extreme conditions of temperature and/or density has become an issue
of great interest in recent years. In this context, it is clearly
important to study how hadron properties (masses, mixing angles, decay
constants, etc.) get modified when hadrons propagate in a hot and/or dense
medium. In particular, since the origin of the light scalar and
pseudoscalar mesons is related to the phenomenon of chiral symmetry
breaking, the temperature and/or density behavior of their properties is
expected to provide relevant information about a possible chiral symmetry
restoration. Unfortunately, even if a significant progress has been made
on the development of ab initio calculations such as lattice
QCD~\cite{All03,Fod04,Kar03}, these are not yet able to provide a full
understanding of the QCD phase diagram and the related hadron
properties, due to the well-known difficulties of dealing with small
current quark masses and finite chemical potentials. Thus it is important
to develop effective models that show consistency with lattice results and
can be extrapolated into regions not accessible by lattice calculation
techniques. In previous works~\cite{GDS00,GDS02,GDS05,DGS04} the study of
the phase diagram of SU(2) chiral quark models that include nonlocal
interactions~\cite{Rip97} has been undertaken. These theories can be
viewed as nonlocal extensions of the widely studied Nambu$-$Jona-Lasinio
model~\cite{reports}. In fact, nonlocality arises naturally in the context
of several successful approaches to low-energy quark dynamics as, for
example, the instanton liquid model~\cite{Schafer:1996wv} and the
Schwinger-Dyson resummation techniques~\cite{RW94}. Lattice QCD
calculations~\cite{Parappilly:2005ei} also indicate that quark
interactions should act over a certain range in momentum space. Moreover,
several studies~\cite{BB95,BGR02} have shown that nonlocal chiral quark
models provide a satisfactory description of hadron properties at zero
temperature and density. On the other hand, when looking at the
description of the chiral phase transition, it has been noticed that for
zero chemical potential these models lead to a rather low critical
temperature $T_c$ in comparison with lattice results~\cite{GDS00,GDS02}.
However, it has been recently shown that the inclusion of the Polyakov
loop, which can be taken as an order parameter for the deconfinement
transition, leads to a significant increase of the chiral restoration
temperature both in two-flavor \cite{Blaschke:2007np} and three-flavor
\cite{Contrera:2007wu} nonlocal models. The inclusion of the Polyakov
loop has also been considered in the context of NJL-like models, namely
the so-called PNJL models~\cite{Meisinger:1995ih,Fukushima:2003fw,
Megias:2004hj,Ratti:2005jh,Roessner:2006xn}, and the quark-meson model
\cite{Schaefer:2009ui}.

The aim of the present work is to go one step beyond previous analyses,
studying the finite temperature behavior of light scalar and pseudoscalar
meson properties in the context of three-flavor nonlocal chiral models
that include mixing with active strangeness degrees of freedom, and taking
care of the effect of gauge interactions by coupling the quarks with a
background color gauge field.

This article in organized as follows. In Sect.\ II we present the general
formalism and derive the expressions needed to evaluate the different
meson properties at finite temperature. In Sect.\ III we provide details
concerning the determination of model parameter values as well as the
results obtained at zero temperature. Our results for the behavior of the
different meson properties as a function of the temperature are presented
and discussed in Sect.\ IV. Finally, in Sect.\ V we sketch our
conclusions.

\section{The formalism}

We deal here with a nonlocal covariant SU(3) quark model which includes
the coupling to a background color gauge field. The Euclidean
effective action for the quark sector of this model is given by
\begin{eqnarray}
S_E &=& \int d^4x \ \left\{ \bar \psi (x) \left[ -i \gamma_\mu D_\mu +
\hat m \right] \psi(x) - \frac{G}{2} \left[ j_a^S(x) \ j_a^S(x) + j_a^P(x)
\ j_a^P(x) \right] \right. \nonumber \\ & & \qquad \qquad \left. -
\frac{H}{4} \ T_{abc} \left[ j_a^S(x) j_b^S(x) j_c^S(x) - 3\ j_a^S(x)
j_b^P(x) j_c^P(x) \right] \ + \ {\cal U}\,[A(x)] \right\}\ ,
\label{se}
\end{eqnarray}
where the chiral U(3) vector $\psi$ includes the light quark fields,
$\psi \equiv (u\; d\; s)^T$, and $\hat m = {\rm diag}(m_u, m_d, m_s)$
stands for the current quark mass matrix. For simplicity we consider the
isospin symmetry limit, in which $m_u = m_d=\bar m$. The fermion kinetic
term includes a covariant derivative $D_\mu\equiv \partial_\mu - iA_\mu$,
where $A_\mu$ are color gauge fields, and the operator
$\gamma_\mu\partial_\mu$ in Euclidean space is defined as $\sum_{i=1,3}
\gamma_i \frac{\partial}{\partial x_i} + \gamma_4\frac{\partial}{\partial
\tau}$, with $\gamma_4=i\gamma_0$. Regarding the interaction terms, the
currents $j_a^{S,P}(x)$ are given by
\begin{eqnarray}
j_a^S (x) & = & \int d^4 z \ \tilde g(z) \
\bar{\psi}\left(x+\frac{z}{2}\right)\ \lambda_a \ \psi
\left(x-\frac{z}{2}\right) \\
j_a^P (x) & = & \int d^4 z \ \tilde g(z) \
\ \bar{\psi}\left(x+\frac{z}{2}\right)\ i\gamma_5 \lambda_a \ \psi
\left(x-\frac{z}{2}\right)\ ,
\end{eqnarray}
where $\tilde g(z)$ is a form factor responsible for the nonlocal
character of the interaction, and the matrices $\lambda_a$, with
$a=0,..,8$, are the standard eight Gell-Mann $3\times 3$ matrices
---generators of SU(3)--- plus $\lambda_0=\sqrt{2/3}\;\openone_{3\times
3}$. The constants $T_{abc}$ in the t'Hooft term (responsible for
flavor-mixing) are defined by
\begin{equation}
T_{abc} = \frac{1}{3!} \ \epsilon_{ijk} \ \epsilon_{mnl} \
\left(\lambda_a\right)_{im} \left(\lambda_b\right)_{jn}
\left(\lambda_c\right)_{kl}\;.
\end{equation}
Finally, the action (\ref{se}) also includes an effective potential ${\cal
U}$ that accounts for gauge field self-interactions.

The partition function associated with the effective action Eq.~(\ref{se})
can be bosonized in the usual way introducing the scalar and pseudoscalar
meson fields $\sigma_a(x)$ and $\pi_a(x)$ respectively, together with
auxiliary fields $S_a(x)$ and $P_a(x)$. To deal with these auxiliary fields
we follow the standard stationary phase approximation, which provides a set
of equations that relate them to the scalar and pseudoscalar meson fields.
Since we are interested in studying the behavior of various meson properties
in the presence of a heat bath, we have to extend the bosonized
effective action to finite temperature. In the present work this is done by
using the Matsubara formalism.

The coupling of fermions to the color gauge fields is implemented
through the covariant derivative in the fermion kinetic term $\gamma_\mu
D_\mu$. As usual, we will assume that the quarks move in a constant
background field $A_4 = i A_0 = i g\,\delta_{\mu 0}\, G^\mu_a
\lambda^a/2$, where $G^\mu_a$ are the SU(3) color gauge fields. Then the
traced Polyakov loop, which is taken as order parameter of confinement, is
given by $\Phi=\frac{1}{3} {\rm Tr}\, \exp( i\beta \phi)$, where $\beta =
1/T$, $\phi = i A_0$. We will work in the so-called Polyakov gauge, in
which the matrix $\phi$ is given a diagonal representation $\phi = \phi_3
\lambda_3 + \phi_8 \lambda_8$, which leaves only two independent
variables, $\phi_3$ and $\phi_8$.

To treat the resulting finite temperature system of interacting mesons in
the presence of the Polyakov loop we consider first the mean field
approximation (MFA), keeping only the nonzero vacuum expectation values
$\bar\sigma_a$. Note that due to charge conservation only $\bar
\sigma_{a=0,3,8}$ can be different from zero. Moreover, $\bar\sigma_3$
also vanishes in the isospin limit. The corresponding MFA grand canonical
thermodynamical potential reads
\begin{eqnarray}
\Omega_{\rm MFA}(T) & = &  -\, 2\,  \sum_{f,c}  \int_{p,n} \mbox{ Tr
ln}\left[ p_{nc}^2 + \Sigma_{f}^2(p_{nc})\right] \; \nonumber \\
& & \qquad \qquad
 -\; \frac{1}{2}\left[ \sum_f (\bar \sigma_f \ \bar S_f  + \frac{G}{2}
\ \bar S_f^2) \; + \; \frac{H}{2} \, \bar S_u\ \bar S_d\ \bar S_s
\right]
 + \; {\cal{U}}(\Phi ,T) \ , \label{ommfa}
\end{eqnarray}
where $f=u,d,s$, $c=r,g,b$, and the shorthand notation $\int_{p,n} = \sum_n
\int d^3p/(2\pi)^3$ has been used. We have also introduced the definition
$p_{nc} = (\vec p\ , \omega_n - \phi_c)$, where $\omega_n$ stands for the
fermionic Matsubara frequencies and the quantities $\phi_c$ are defined by
the relation $\phi = {\rm diag}(\phi_r,\phi_g,\phi_b)$. The quark
constituent masses $\Sigma_{f}(p_{nc})$ are here momentum-dependent
quantities, given by
\begin{equation}
\Sigma_{f}(p_{nc}) \ = \ m_f\, + \, \bar\sigma_f\, g(p_{nc})\ ,
\label{sigma}
\end{equation}
where $g(p)$ is the Fourier transform of the form factor $\tilde g(z)$.
For convenience we have introduced mean field values $\bar\sigma_f$ given
by
\begin{eqnarray}
\bar \sigma_u & = & \sqrt{\frac23} \, \bar \sigma_0 + \bar \sigma_3
+ \frac1{\sqrt{3}} \, \bar \sigma_8 \nonumber \;
, \quad
\bar \sigma_d  =  \sqrt{\frac23} \, \bar \sigma_0 - \bar \sigma_3 +
\frac1{\sqrt{3}} \, \bar \sigma_8 \nonumber \; , \quad \bar \sigma_s  =
\sqrt{\frac23} \, \bar \sigma_0 + \frac2{\sqrt{3}} \, \bar \sigma_8\;,
\end{eqnarray}
and similar definitions hold for $\bar S_f$ in terms of $\bar S_0$, $\bar
S_3$ and $\bar S_8$. Note that in the isospin limit $\bar \sigma_u = \bar
\sigma_d$, thus we have $\Sigma_u(p_{nc}) = \Sigma_d(p_{nc})$. Within the
stationary phase approximation, the mean field values of the auxiliary
fields $\bar S_f$ turn out to be related with the mean field values of the
scalar fields $\bar \sigma_f$ by~\cite{Scarpettini:2003fj}
\begin{equation}
\bar \sigma_u + G\,\bar S_u + \frac{H}{2} \, \bar S_u \bar S_s =
0\ \ , \qquad \bar \sigma_s + G\,\bar S_s + \frac{H}{2} \, \bar
S_u^2 = 0\ \ . \label{gapeq}
\end{equation}

The effective potential ${\cal{U}}(\Phi ,T)$, which accounts for Polyakov
loop dynamics, can be fitted by taking into account group theory
constraints together with lattice results, from which one can estimate the
temperature dependence. Following Ref.~\cite{Roessner:2006xn} we take
\begin{equation}
{\cal{U}}(\Phi ,T) = \left[-\,\frac{1}{2}\, a(T)\,\Phi^2 \;+\;b(T)\, \ln(1
- 6\, \Phi^2 + 8\, \Phi^3 - 3\, \Phi^4)\right] T^4 \ ,
\label{effpot}
\end{equation}
with the corresponding definitions of $a(T)$ and $b(T)$. Owing to the
charge conjugation properties of the QCD Lagrangian~\cite{Dumitru:2005ng},
the mean field traced Polyakov loop field $\Phi$ is expected to be a real
quantity. Assuming that $\phi_3$ and $\phi_8$ are
real-valued~\cite{Roessner:2006xn}, this implies $\phi_8 = 0$, $\Phi = [ 2
\cos(\phi_3/T) + 1 ]/3$.

For finite current quark masses the quark contribution to
$\Omega_{\rm MFA}(T)$ turns out to be divergent. To regularize it
we follow the same prescription as in previous works~\cite{GDS05}.
Namely, we subtract from $\Omega_{\rm MFA}(T)$ the quark
contribution in the absence of fermion interactions, and then we
add it in a regularized form, i.e.\ after the subtraction of an
infinite, $T$-independent contribution. From the minimization of
this regularized thermodynamical potential it is possible now to
obtain a set of three coupled ``gap'' equations that determine the
mean field values $\bar\sigma_u$, $\bar\sigma_s$ and $\phi_3$
at a given temperature:
\begin{equation}
\frac{\partial \Omega_{\rm MFA} } {\left( \partial\bar\sigma_u,
\partial\bar\sigma_s, \partial\phi_3 \right)} \ = \ 0\ . \label{gap}
\end{equation}

In order to obtain the meson mass spectrum and other properties one has to
consider the mesonic fluctuations around the mean field values. We begin by
introducing a more convenient basis defined by
\begin{equation}
\xi_{ij} = \frac{1}{\sqrt2}\,\left(\lambda_a\ \xi_a\right)_{ij}\;,
\end{equation}
where $\xi_a=\sigma_a,\pi_a$, while $i,j$ run from 1 to 3 (neutral fields
are shifted by $\xi_a \to\xi_a - \bar\xi$). For the scalar fields one has in
this way
\begin{equation}
\sigma_{ij} = \left( \begin{array}{ccc} \displaystyle
\frac{a_0^0}{\sqrt{2}} + \frac{\sigma_8}{\sqrt{6}} +
\frac{\sigma_0}{\sqrt3} & a_0^+ & \kappa^+ \\ a_0^- & \displaystyle -
\frac{a_0^0}{\sqrt{2}} + \frac{\sigma_8}{\sqrt{6}} +
\frac{\sigma_0}{\sqrt3} & \kappa^0 \\ \kappa^- & \bar \kappa^0 & \displaystyle -
\frac{2\,\sigma_8}{\sqrt{6}} + \frac{\sigma_0}{\sqrt3}
\end{array} \right)_{ij}\ ,
\end{equation}
while a similar expression holds for the pseudoscalar sector, replacing
$a_0\to\pi$, $\sigma \to\eta$ and $\kappa\to K$. Using this notation the
resulting quadratic contribution to the finite temperature bosonized
effective action can be written as
\begin{equation}
S_E^{\rm quad} = \frac{1}{2} \int_{q,m} \left[ G^+_{ij,kl}(\vec q \ ^2
, \nu_m^2) \ \sigma_{ij}(q_m) \ \sigma_{kl}(-q_m) +
G^-_{ij,kl}(\vec q \ ^2 , \nu_m^2) \
\pi_{ij}(q_m)\ \pi_{kl}(-q_m) \right]\;,
\label{sequad}
\end{equation}
where $q_m=(\vec q\ , \nu_m )$, $\nu_m = 2 m \pi T$ being bosonic
Matsubara frequencies. The functions $G^\pm_{ij,kl}$ in Eq.~(\ref{sequad})
are given by
\begin{equation}
G^{\pm}_{ij,kl}(\vec q \ ^2 , \nu_m^2) = C^{\pm}_{ij}(\vec q \ ^2
, \nu_m^2) \ \delta_{il}\,\delta_{jk} +
\left((r^\pm)^{-1}\right)_{ij,kl}\;,
\end{equation}
where
\begin{eqnarray}
C^\pm_{ij}(\vec q \ ^2 , \nu_m^2)
& = &
 -\, 8 \,\sum_c \, \int_{p,n} \ g(p_{nc}+q_m/2)^2\
\frac{p_{nc}^2 + p_{nc} \cdot q_m \mp \Sigma_i(p_{nc}+q_m) \Sigma_j(p_{nc})}
{D_i(p_{nc}+q_m)  D_j(p_{nc})}\ ,
\label{ciju}
\end{eqnarray}
and
\begin{eqnarray}
r^\pm_{ij,kl} & = & G  \ \delta_{il}\,\delta_{jk} \pm \frac{H}{2}\
\epsilon_{ikh}\,\epsilon_{jlh} \ \bar S_h \ .
\end{eqnarray}
In Eq.~(\ref{ciju}) we have defined $D_j (s) \equiv s^2 + \Sigma_j^2(s)$,
where $j=1,2,3$ correspond to $f=u,d,s$ in the notation of
Eq.~(\ref{sigma}).

{}From the quadratic effective action $S_E^{\rm quad}$ it is possible to
derive the scalar and pseudoscalar meson masses as well as the quark-meson
couplings. In what follows we will consider explicitly only the case of
pseudoscalar mesons. The corresponding expressions for the scalar sector are
completely equivalent, just replacing upper indices ``$-$'' by ``$+$''. In
terms of the physical fields, the contribution of pseudoscalar mesons to
$S_E^{\rm quad}$ can be written as
\begin{eqnarray}
\left. S_E^{\rm quad}\right|_P & = & \frac12  \int_{q,m} \, \bigg\{
G_\pi (\vec q \ ^2 , \nu_m^2) \left[ \pi^0 (q_m)\ \pi^0 (-q_m) +
2\,\pi^+ (q_m) \ \pi^- (-q_m)\right]
\nonumber \\
& & \qquad \qquad + G_K (\vec q \ ^2 , \nu_m^2) \left[ 2\, K^0
(q_m) \ \bar K^0 (-q_m) + 2\, K^+ (q_m) \ K^- (-q_m) \right]
\nonumber \\
& & \qquad \qquad + G_\eta(\vec q \ ^2 , \nu_m^2) \ \eta(q_m) \
\eta(-q_m)+ G_{\eta'}(\vec q \ ^2 , \nu_m^2) \ \eta'(q_m) \ \eta'(-q_m) \bigg\}\;.
\label{quad}
\end{eqnarray}
Here, the fields $\eta$ and $\eta'$ are related to the U(3) states $\eta_0$
and $\eta_8$ according to
\begin{eqnarray}
\eta &=& \cos \theta_\eta \ \eta_8 - \sin \theta_\eta\ \eta_0
\nonumber \\
\eta' &=& \sin \theta_{\eta'} \ \eta_8 + \cos \theta_{\eta'}\ \eta_0 \ ,
\label{mixing}
\end{eqnarray}
where the mixing angles $\theta_{\eta,\eta'}$ are defined in such a way that
there is no $\eta-\eta'$ mixing at the level of the quadratic action. In a
similar way, in the scalar sector one has two physical scalar mesons
$\sigma$ and $f_0(980)$ that are linear combinations of the states
$\sigma_8$ and $\sigma_0$, with mixing angles $\theta_\sigma$ and
$\theta_{f_0}$. The functions $G_P(\vec q \ ^2 , \nu_m^2)$ introduced in
Eq.~(\ref{quad}) are given by
\bea
G_\pi(\vec q \ ^2 , \nu_m^2) \! & \! = \! & \!
\left[
(G + \frac{H}{2} \bar S_s)^{-1} + C^-_{uu}(\vec q \ ^2 , \nu_m^2)
\right] \label{gpi} \nonumber \\
G_K (\vec q \ ^2 , \nu_m^2) \! & \! = \! & \!
\left[ (G + \frac{H}{2} \bar
S_u)^{-1} + C^-_{us}(\vec q \ ^2 , \nu_m^2)
\right] \\
G_{\eta} (\vec q \ ^2 , \nu_m^2) \! & \! = \! & \!
\frac{G^-_{88}(\vec q \ ^2
, \nu_m^2) + G^-_{00}(\vec q \ ^2 , \nu_m^2)}{2} - \sqrt{ \left[
G^-_{08}(\vec q \ ^2 , \nu_m^2) \right]^2 \! + \! \left[
\frac{G^-_{88}(\vec q \ ^2 , \nu_m^2) -
G^-_{00}(\vec q \ ^2 , \nu_m^2)}{2} \right]^2} \nonumber \\
G_{\eta'} (\vec q \ ^2 , \nu_m^2) \! & \! = \! & \!
\frac{G^-_{88}(\vec q \ ^2 , \nu_m^2) + G^-_{00}(\vec q \ ^2 ,
\nu_m^2)}{2} + \sqrt{ \left[ G^-_{08}(\vec q \ ^2 , \nu_m^2)
\right]^2 \! + \! \left[ \frac{G^-_{88}(\vec q \ ^2 , \nu_m^2) -
G^-_{00}(\vec q \ ^2 , \nu_m^2)}{2} \right]^2}\;, \nonumber \eea
where \bea G^-_{88} (\vec q \ ^2 , \nu_m^2)  & =  & \frac13
\frac{6 G - 4 H \bar S_u - 2 H \bar S_s}
               {2 G^2 - G H \bar S_s - H^2 \bar S_u^2} +
C^-_{88} (\vec q \ ^2 , \nu_m^2) \nonumber
\\
G^-_{08} (\vec q \ ^2 , \nu_m^2)  &  =  & \frac{\sqrt{2}}{3}
\frac{H (\bar S_s - \bar S_u)}
                     {2 G^2 - G H \bar S_s - H^2 \bar S_u^2} +
C^-_{08} (\vec q \ ^2 , \nu_m^2)
\label{g00}
\\
G^-_{00} (\vec q \ ^2 , \nu_m^2)   &  =   & \frac13 \frac{6 G + 4
H \bar S_u - H \bar S_s}
               {2 G^2 - G H \bar S_s - H^2 \bar S_u^2}
+  C^-_{00} (\vec q \ ^2 , \nu_m^2) \nonumber \,.
\eea
and
\bea
C^-_{88} = \frac{C^-_{uu} + 2\,C^-_{ss}}{3}
\ , \qquad
C^-_{08} = \frac{\sqrt{2}}{3} \left(C^-_{uu} - C^-_{ss}\right)
\ , \qquad
C^-_{00} = \frac{2\,C^-_{uu} + C^-_{ss}}{3}\ .
\eea

Now the pseudoscalar meson masses are obtained by solving the equations
\begin{equation}
G_P (- m_P^2,  0) = 0\;,
\label{gp}
\end{equation}
with $P=\pi$, $K$, $\eta$ and $\eta'$. The mass values determined by these
equations correspond to the spatial ``screening-masses'' of the mesons'
zeroth Matsubara modes, and their inverses describe the persistence lengths
of these modes at equilibrium with the heat bath. It is worth to notice that
there is a screening mass for each Matsubara mode. The full bound state
propagator can be calculated via any polarization tensor that receives a
contribution from the bound state, but only once all screening masses
have been determined. The propagator obtained in this way is defined only on
a discrete set of points along what might be called the imaginary-energy
axis, and the ``pole-mass'', i.e., the mass that yields the bound state
energy pole for $\vec q \sim 0$, is obtained only after an analytic
continuation of the propagator onto the real-energy axis. The fact that
Lorentz invariance is broken for $T > 0$ means that, in general, the pole
mass and screening masses are not equal (see e.g.\
Ref.~\cite{Florkowski:1993br}). Although the analytic continuation involved
in this process is not unique, an unambiguous result is obtained by
requiring that it yields a function that is bounded at complex-infinity and
analytic off the real axis~\cite{LW87}. From this description it is
nonetheless clear that the screening masses completely specify the
properties of $T>0$ bound states. The masses associated to the zeroth
Matsubara mode studied here are spatial screening masses corresponding to a
behavior $\exp(-m_P \, r)$ in the conjugate 3-space coordinate $r$, and
should correspond to the lowest state in each meson channel. In fact,
these are the quantities usually studied in lattice calculations
\cite{Karsch:2003jg}.

In the $\eta-\eta'$ system, once the meson masses have been determined one
can find the mixing angles $\theta_\eta$ and $\theta_{\eta'}$, which are in
general different from each other. These are given by
\begin{eqnarray}
\mbox{tan}\ 2\,\theta_P &=& \frac{2\,
G^-_{08}(-m^2_P,0)}{G^-_{00}(-m^2_P,0) -
G^-_{88}(-m^2_P,0)} \ \ , \qquad P \ =\ \eta,\ \eta' \ .
\label{thetap}
\end{eqnarray}

The meson fields have to be renormalized, so that the residues of the
corresponding propagators at the meson poles are set equal to one. The
corresponding wave function renormalization constants $Z_P$ are given by
\begin{equation}
Z_P^{-1}  = \frac{d G_P(\vec q\ ^2,0) }{d \vec q\ \! ^2} \bigg|_{\ \vec q\
\! ^2=-m_P^2}\;, \label{zp1}
\end{equation}
with $P=\pi$, $K$, $\eta$ and $\eta'$. Finally, the meson-quark
coupling constants $G_{Pq}$ are given by the original residues of
the meson propagators at the corresponding poles,
\begin{equation}
G_{Pq} = Z^{1/2}_P\; .
\label{gpq}
\end{equation}

In the case of the pseudoscalar mesons other important features are the
corresponding weak decay constants $f_{ab}$, defined by
\begin{equation}
\langle\, 0 | A^a_\mu(0) | \pi_b (q) \,\rangle = i
\;f_{ab}\; q_\mu\;.
\end{equation}
where $A^a_\mu$ is the $a$-component of the axial current. For $a,b=1\dots
7$, the constants $f_{ab}$ can be written as $\delta_{ab}\,f_P$, with $P =
\pi$ for $a=1,2,3$ and $P=K$ for $a=4$ to 7. In contrast, as occurs with the
meson masses, the decay constants become mixed in the $a=0,8$ sector.
Details on how to obtain the expressions for the axial currents in the
presence of nonlocal fields can be found e.g.\ in
Refs.~\cite{BB95,Scarpettini:2003fj}. After a rather lengthy calculation we
find that, at finite temperature, the pion and kaon decay constants are
given by \bea
f_\pi &=& 4\, f_{uu}(-m_\pi^2, 0)\  Z_\pi^{1/2}\;, \label{fpi}
\nonumber \\
f_K   &=& 2 \left[ f_{us}(-m_K^2, 0) + f_{su}(-m_K^2, 0) \right]
Z_K^{1/2}\;, \label{fk}
\eea
where
\begin{eqnarray}
f_{ij}(\vec q\ \!^2,\nu_m) & = &
\sum_c \, \int_{p,n}
\Bigg\{
g\left(p_{nc}+\frac{q_m}{2}\right)
\frac{ p_{nc} \cdot q_m \Sigma_i(p_{nc}+ q_m) - ( p_{nc} \cdot q_m + p_{nc}^2 ) \Sigma_j(p_{nc}) }
{D_i(p_{nc}+q_m)  D_j(p_{nc})}
\nonumber \\
& & \qquad  -
\left[ 2 g\left(p_{nc}+\frac{q_m}{2}\right) - g\left(p_{nc}\right) - g\left(p_{nc}+ q_m\right) \right]
\frac{\Sigma_i(p_{nc}+ \frac{q_m}{2})}{D_i(p_{nc}+\frac{q_m}{2})}
\nonumber \\
& & \qquad
+ \ g\left(p_{nc}+\frac{q_m}{2}\right)
\left[ \Sigma_i\left(p_{nc}+\frac{q_m}{2}\right) + \Sigma_j\left(p_{nc}+\frac{q_m}{2}\right) -
\Sigma_i\left(p_{nc}\right)-\Sigma_j\left(p_{nc}+ q_m \right) \right]
\nonumber \\
& & \qquad \qquad  \times\;
\frac{p_{nc}^2 + p_{nc} \cdot q_m + \Sigma_i(p_{nc}+q_m) \Sigma_j(p_{nc})}
{D_i(p_{nc}+q_m)  D_j(p_{nc})}\Bigg\}\;.
\label{fij}
\label{expfpi}
\end{eqnarray}

In the case of the $\eta-\eta'$ system, two decay constants can be defined
for each component ($a=0$ or 8) of the axial current. They can be written
in terms of the decay constants $f_{ab}$ and the previously defined mixing
angles $\theta_{\eta,\eta'}$ as
\begin{eqnarray}
f^a_{\eta} &=& \left[ f_{a8}(-m_{\eta}^2,0) \cos \theta_{\eta} -
f_{a0}(-m_{\eta}^2,0) \sin \theta_{\eta}\right] \ Z_\eta^{1/2}
\nonumber \\
f^a_{\eta'} &=& \left[ f_{a8}(-m_{\eta'}^2,0) \sin \theta_{\eta'} +
f_{a0}(-m_{\eta'}^2,0) \cos \theta_{\eta'} \right] \
Z_{\eta'}^{1/2}\;.
\label{faetas}
\end{eqnarray}
Within our model, the decay constants $f_{ab}$ for $a,b=0,8$ are related
to the $f_{ij}$ defined in Eq.~(\ref{fij}) by
\bea
f_{88}(\vec q\ \!^2,\nu_m)  &=&
\frac{4}{3} \left[ 2 f_{ss}(\vec q\ \!^2,\nu_m) + f_{uu}(\vec q\ \!^2,\nu_m) \right] \nonumber \\
f_{00}(\vec q\ \!^2,\nu_m)  &=& \frac{4}{3} \left[ 2 f_{uu}(\vec q\ \!^2,\nu_m) + f_{ss}(\vec q\ \!^2,\nu_m) \right] \\
f_{08}(\vec q\ \!^2,\nu_m) = f_{80}(\vec q\ \!^2,\nu_m) &=&
\frac{4\sqrt{2}}{3} \left[ f_{uu}(\vec q\ \!^2,\nu_m) - f_{ss}(\vec q\
\!^2,\nu_m) \right]\;. \label{f08} \nonumber
\eea
As expected, both the nondiagonal decay constants $f_{08}$, $f_{80}$ and the
mixing angles $\theta_\eta$, $\theta_{\eta'}$ vanish in the SU(3) symmetry
limit.

\section{Model parameters and zero temperature results}

In this section we determine the model parameters to be used in our
numerical calculations, and quote the results obtained for various meson
properties at zero temperature. The latter include the values of meson
masses, decay constants and mixing angles, as well as quark constituent
masses, quark condensates and quark-meson couplings.

At low temperatures, the value of the traced Polyakov loop is essentially
determined by the effective potential in Eq.~(\ref{effpot}), therefore for
$T\to 0$ one has $\Phi\to 0$, $\cos(\phi_3/T)\to -1/2$. Since for low $T$
the Matsubara sums in the thermodynamical potential are governed by modes
with large $n$, one has $\omega_n - \phi_c = [(2n+1)\pi -\phi_c/T]T\simeq
\omega_n$, thus for $T\to 0$ the coupling of fermions to the Polyakov loop
vanishes. In this way, the zero-$T$ calculations are similar to those
carried out in Ref.~\cite{Scarpettini:2003fj}, in which SU(3) nonlocal
chiral quark models without the inclusion of the Polyakov loop have been
considered. As in that work, our numerical analysis has been performed using
a Gaussian form factor, namely
\begin{equation}
g(p) \ = \ \exp{\left(-p^2/\Lambda^2\right)} \ ,
\label{formf}
\end{equation}
which has been often considered in the literature. This introduces a new
free parameter $\Lambda$, which plays the r\^ole of an ultraviolet cut-off
momentum scale (we recall that the form factor is defined in Euclidean
momentum space). At $T=0$, the main difference between our analysis and that
in Ref.~\cite{Scarpettini:2003fj} is that here we are considering an
OGE-motivated nonlocal interaction, whereas in the previous work a different
form [motivated by instanton liquid models (ILM)] for the nonlocal currents
has been chosen. In the case of two-flavor models, a detailed
comparison between these different interaction forms has been carried out in
Ref.~\cite{GomezDumm:2006vz}, showing that the results for both models are
qualitatively similar. Notice that in Ref.~\cite{Scarpettini:2003fj} only
the pseudoscalar meson sector was addressed.

After the assumption of the form factor in Eq.~(\ref{formf}), the nonlocal
chiral quark model under consideration includes five free parameters, namely
the current quark masses $\bar m$ and $m_s$, the coupling constants $G$ and
$H$ and the cut-off scale $\Lambda$. In our numerical calculations we have
chosen to fix the value of $\bar m$, whereas the remaining four parameters
are determined by requiring that the model reproduces correctly the measured
values of four physical quantities at zero temperature. These are the masses
of the pion, kaon and $\eta'$ pseudoscalar mesons, and the pion decay
constant $f_\pi$. Taking $\bar m = 5$ MeV, we obtain the following set of
parameters:
\begin{eqnarray}
\bar m & = & 5 \ {\rm MeV}\ \ {\rm (input)} \nonumber \\
m_s & = & 119 \ {\rm MeV} \nonumber \\
\Lambda & = & 843 \ {\rm MeV} \nonumber \\
G\Lambda^2 & = & 13.35 \nonumber \\
H\Lambda^5 & = & - 273.7 \ \ % .
\label{param}
\end{eqnarray}

Our numerical results are presented in Table I. For comparison, in the last
column of this table we quote the measured values of meson masses, and the
ranges in which the decay constants and mixing angles should fall according
to most popular phenomenological approaches. Entries marked with an asterisk
are those that we have taken as input values to fix the model parameters.
From Table I it is seen that in general there is a reasonable agreement
between the predicted meson masses and the empirical values quoted by the
Review of Particle Physics~\cite{PDG2008}. In addition, the obtained mass
ratio $m_s/m = 23.8$ is close to the corresponding current algebra
prediction, namely $m_s/m = (2m_K^2- m_\pi^2)/m_\pi^2 \simeq 25$. We point
out that in the case of the mass of the $\kappa$ scalar meson the equation
$G_\kappa(-x^2,0)=0$ has no solution in the real $x$ axis. Hence we have
defined the mass $m_\kappa$ as the point where the absolute value of
$G_\kappa(-x^2,0)$ becomes minimal. A more sophisticated definition could be
done by extending $x$ to the complex plane, thus introducing a finite
$\kappa$ width. In any case, this would not change significantly the mass
value. A detailed analysis of the regularization prescriptions for the
evaluation of loop integrals like those in Eqs.~(\ref{ciju}) and (\ref{fij})
has been carried out in Ref.~\cite{Scarpettini:2003fj}.

\begin{table}
\begin{tabular}{cccccccccc} \hline
\hline
                    &           & Our Model  && $\begin{array}{c} {\rm Empirical \ \&} \\
                                                {\rm   Phenomenological} \end{array}$  \\
\hline
$\bar m$            &  [ MeV ]  & $5^*$        && (3.4\ -\ 7.4)    \\
$m_s$               &  [ MeV ]  & 119        && (108\ -\ 209)    \\
\hline
$m_\pi$             &  [ MeV ] &  $139^*$    &&  139  \\
$m_K$               &  [ MeV ] &  $495^*$    &&  495  \\
$m_\eta$            &  [ MeV ] &   523\;\,   &&  547  \\
$m_{\eta'}$         &  [ MeV ] &  $958^*$    &&  958  \\
\hline
$m_{a_0}$           &  [ MeV ] & 900           &&  980  \\
$m_\kappa$          &  [ MeV ] & 1380          &&  1425  \\
$m_\sigma$          &  [ MeV ] & 566           &&  400-1200  \\
$m_{f_0}$           &  [ MeV ] & 1280          &&  980  \\
\hline
$G_{\pi q}$         &&  3.98  &     \\
$G_{K q}$           &&  4.30  &     \\
$G_{\eta q}$        &&  3.93  &     \\
$G_{\eta' q}$       &&  2.83  &     \\
\hline
$\theta_\eta$    && $-2.3^\circ$  &&  \\
$\theta_{\eta'}$ && $-40.3^\circ$ &&  \\
\hline
$\theta_8$       && $-24^\circ$ && $-$($22^\circ$\ - $19^\circ$) \\
$\theta_0$       && $-7.7^\circ$  && $-$($10^\circ$\ - $0^\circ$)  \\
\hline
$f_\pi$                 &  [ MeV ] &  $92.4^*$\!\!\!   && 92.4         \\
$f_K/f_\pi$             &          &   1.17      && 1.22         \\
$f_{\eta}^{8}/f_\pi$    &          &   1.14      && (1.17-1.22)  \\
$f_{\eta} ^{0}/f_\pi$   &          &   0.16      && (0.11-0.37)  \\
$f_{\eta '} ^{8}/f_\pi$ &          &  -0.49      && -(0.42-0.46) \\
$f_{\eta '} ^{0}/f_\pi$ &          &   1.16      && (0.98-1.16)  \\
\hline
(*) Input values & & & &
\end{tabular}
\caption{$T=0$ model predictions for various meson properties: masses,
mixing angles, decay constants and quark-meson couplings.}
\end{table}

Concerning the pseudoscalar meson decay constants, we notice that the
predicted value for $f_K$ is also phenomenologically acceptable. In fact,
it turns out to be significantly better than that obtained in the standard
NJL model, where the kaon and pion decay constants are found to be
approximately equal to each other~\cite{reports} in contrast with
experimental evidence. Regarding the mixing angles and decay constants for
the $\eta_8-\eta_0$ system, the problem of defining and (indirectly)
fitting these parameters has been revisited several times in the
literature (see e.g.\ Ref.~\cite{F00}, and references therein). As stated
in the previous section, in general one has to deal with two different
state mixing angles $\theta_P$ and four decay constants $f_P^a$, where
$P=\eta,\eta'$ and $a=0,8$. This means that $\eta$ and $\eta'$ states do
not need to be orthogonal, and the same occurs with
$(f_\eta^8,f_{\eta'}^8)$ and
$(f_\eta^0,f_{\eta'}^0)$~\cite{L97,FK02,EF99}. For the sake of comparison
with phenomenological values of these parameters, we follow here
Ref.~\cite{L97} and express the four decays constants $f_P^a$ in terms of
two decay constants $f_a$ and two mixing angles $\theta_a$, where $a=0,8$:
\begin{equation}
\left(
\begin{array}{cc}
f_\eta^8 & f_\eta^0 \\ f_{\eta'}^8 & f_{\eta'}^0
\end{array}
\right) =
\left(
\begin{array}{cc}
f_8\,\cos\theta_8 & -f_0\,\sin\theta_0 \\
f_8\,\sin\theta_8 & \ \ f_0\,\cos\theta_0
\end{array}
\right)\ .
\label{fleut}
\end{equation}
In our framework the decay constants $f_P^a$ can be calculated from
Eqs.~(\ref{faetas}). As shown in Table I, both the values obtained for
$\theta_8$, $\theta_0$ as well as those obtained for the decay constants
in the $\eta-\eta'$ sector are in agreement with phenomenological results.
These have been taken from the analysis in Ref.~\cite{F00}, in which the
values obtained from different parameterizations have been translated to
the four-parameter decay constant scheme given by Eq.~(\ref{fleut}).
Notice that $\theta_8$ and $\theta_0$ are significantly different to each
other, as occurs with the mixing angles $\theta_\eta$ and $\theta_{\eta'}$
[which can be calculated from Eq.~(\ref{thetap})]. This is in agreement
with the analysis in Ref.~\cite{L97}, carried out within next-to-leading
order Chiral Perturbation Theory and large $N_C$, which leads to $\theta_8
= -20.5^\circ$, $\theta_0 = -4^\circ$.

\section{Finite temperature results}

Taking the parameters in Eq.~(\ref{param}), one can solve Eqs.~(\ref{gap})
to calculate the mean field values $\bar\sigma_u$, $\bar\sigma_s$ and
$\phi_3$ at finite temperature. The behavior of effective quark masses and
condensates, as well as the curves for the traced Polyakov loop $\Phi$,
are similar to those obtained in Ref.~\cite{Contrera:2007wu} within an
ILM-motivated nonlocal chiral model. The discussion of those results is
qualitatively the same as in our case, therefore it will not be repeated
here. We just state that, as expected, there is a crossover phase
transition in which chiral symmetry is restored, and consequently one
finds a sharp peak in the chiral susceptibility. The transition
temperature (defined as the position of this peak) is found to be $T_c =
202$~MeV. This value is in much better agreement with lattice results,
$T_c^{\rm (latt)} = 160-200$ MeV~\cite{Bernard:2004je}, than the value
recently obtained in the local SU(3) PNJL model, $T^{\rm (PNJL)}_c =
259$~MeV~\cite{Costa:2008dp}. In addition one finds a deconfinement phase
transition, which occurs at about the same critical temperature.

\begin{figure}[htb]
\includegraphics[width=0.59\textwidth]{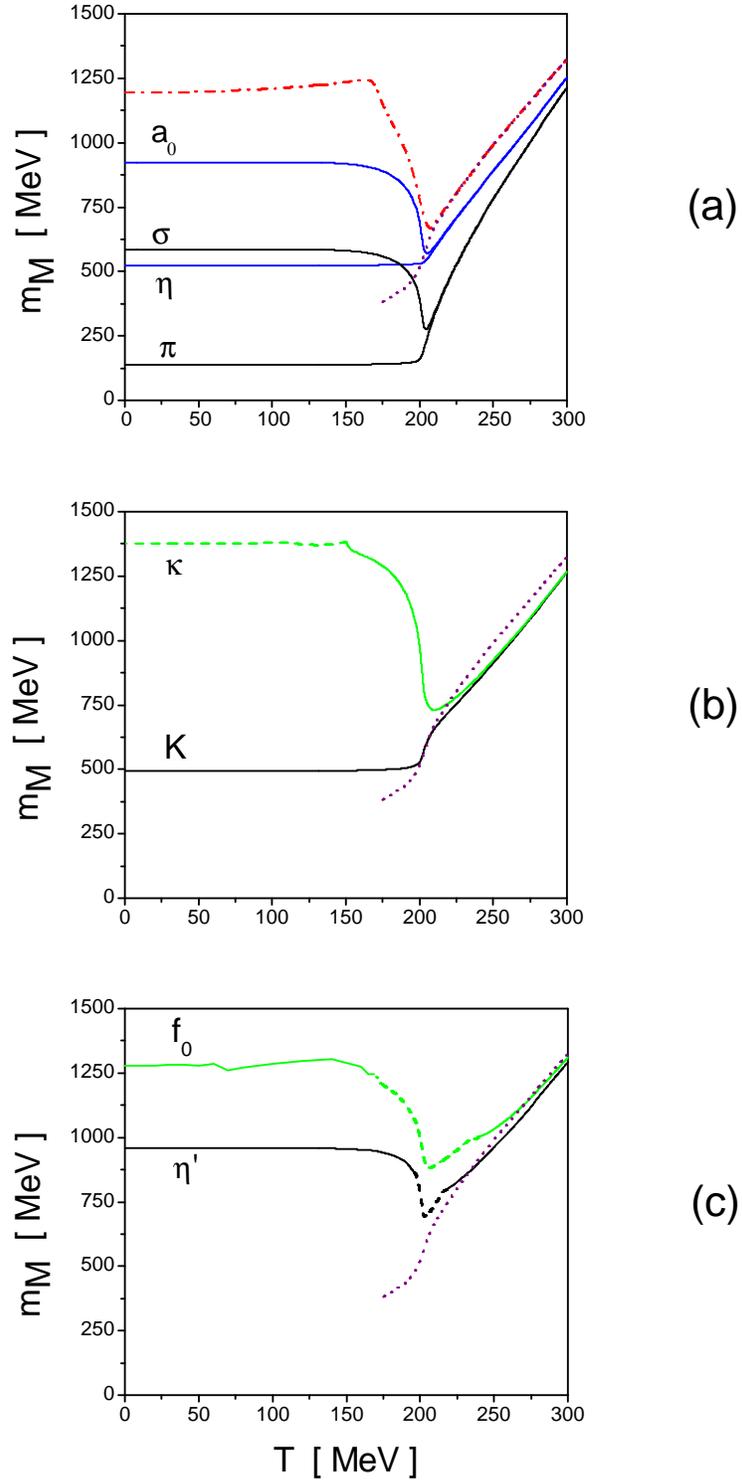}
\caption{Solid lines quote scalar and pseudoscalar meson masses as
functions of the temperature. Dotted and dashed-dotted lines stand for the
value $m_M^{\rm uq} = 2(\pi T - \phi_3)$ and the $\bar qq$
production threshold, respectively (see text).}
\end{figure}

We concentrate here in the evolution of meson masses and decay constants
with temperature, which has not been previously addressed in the context of
SU(3) nonlocal models. Pseudoscalar meson masses can be determined by
solving Eqs.~(\ref{gp}), while the same procedure applies to the scalar
meson sector replacing upper indices ``$-$´´ by ``$+$'' in
Eqs.~(\ref{gpi}-\ref{g00}). As discussed in Sect.~II, these values
correspond to the spatial screening-masses of the mesons' zeroth Matsubara
modes. Our numerical results are shown in Fig.~1, where we quote the values
of meson masses as functions of the temperature. In Fig.~1(a) we show the
behavior of the pseudoscalar mesons $\pi$ and $\eta$ together with the
curves for the scalar mesons $\sigma$ and $a_0$, which are chiral partners
of the former. It is seen that pseudoscalar meson masses remain
approximately constant up to the critical temperature (this is reasonable,
since they are protected from chiral symmetry), while scalar meson masses
begin to drop at about 150~MeV. Above $T_c$ pseudoscalar masses get
increased, in such a way that they become degenerate with the masses of
their chiral partners, as expected from chiral restoration. In particular,
the fact that this occurs right after the transition in the case of the
$(\eta,a_0)$ pair indicates that the strange contents of the $\eta$
and $a_0$ mesons become suppressed above the critical temperature. When the
temperature is further increased, all four masses are found to rise
continuously, showing that now the mass is basically dominated by thermal
energy. At very large temperatures the curves should approach asymptotically
the value corresponding to a $q \bar q$ pair of uncorrelated
massless quarks $m_M^{\rm uq} = 2 \pi T$~\cite{Eletsky:1988an}. At
$T\approx 300$~MeV, however, the Polyakov loop has not yet reached its
asymptotic value $\phi_3/T|_{T\rightarrow \infty}=0$, and it still provides a
non-negligible correction to the quark screening mass. In fact, we find
$\phi_3/T|_{T=300}\simeq 0.93$. Thus, around this temperature we expect
$m_M^{\rm uq} = 2(\pi T - \phi_3)$, which is shown by the
dotted lines in Figs.~1(a), (b) and (c). For vanishing quark dynamical masses, this value of
$m_M$ corresponds to a pole in the $n=0$ mode for the integrals in the
functions $C_{ii}^\pm(-k^2,0)$, see Eq.~(\ref{ciju}). Indeed, as discussed
in Ref.~\cite{Blaschke:2000gd}, Eqs.~(\ref{gp}) can be satisfied only in the
vicinities of these poles. On the other hand, in general it is seen that the
functions $C_{ij}^\pm(-k^2,0)$ [and therefore also the functions
$G_M(-k^2,0)$] are well defined for low values of $k$. If $k$ is increased,
at some critical point $k_{\rm crit}$, usually called ``pinch point'', the
integrals become divergent and need some regularization prescription. In the
present work we follow the prescription discussed in the Appendix of
Ref.~\cite{Scarpettini:2003fj}, conveniently extended to the finite
temperature case. The pinch point occurs when both effective quarks are
simultaneously on-shell, thus it can be interpreted as a threshold above
which mesons could decay into two massive quarks. In Fig.~1(a) this
threshold is represented with the dashed-dotted curve (above $T_c$, it
approximately matches the value of $m_M^{\rm uq}$ mentioned previously).
It can be seen that all four meson masses in Fig.~1(a) remain below the
threshold for the temperature range considered.

In Fig.~1(b) we represent the curves for the masses of the pseudoscalar
mesons $K$, and their scalar partners $\kappa$. It is seen that for some
temperature range the equation $G_\kappa(-k^2,0)=0$ has no solution for
real $k$, therefore the mass is defined as the minimum of the function
$G_\kappa(-k^2,0)$, as discussed in the previous section. These mass
values correspond to the dashed stretch of the corresponding curve. It is
worth to notice that the $K$ and $\kappa$ meson masses match only at
$T\simeq 225$~MeV, i.e.\ at a temperature somewhat larger than $T_c$. This
is clearly a consequence of the large current strange quark mass, which is
expected to move the SU(3) chiral restoration to higher temperatures.
Finally, in Fig.~1(c) we quote the temperature dependence of f$_0$ and
$\eta'$ masses. As before, dashed stretches in the curves indicate the
regions in which the corresponding function $G_M(-k^2,0)$ has no zero for
real $k$ and, therefore, the mass $m_M$ is defined by the position of its
minimum. Let us first focus on the behavior of the $\eta'$ mass. In
contrast with some results found in Ref.~\cite{Horvatic:2007qs}, where
the corresponding temperature dependence has been studied in the framework
of a Dyson-Schwinger approach, we do not observe any kind of enhancement
of $m_{\eta'}$ around $T_c$. It should be noticed that in the framework of
Ref.~\cite{Horvatic:2007qs} the effect of the U(1)$_A$ anomaly is modelled
in a simpler way, namely by considering it only at the level of mass
shifts. In this sense our result is consistent with the analyses of the
$\eta'$ pole mass performed within the local SU(3) PNJL model~\cite{Costa:2008dp}
and the quark-meson model~\cite{Schaefer:2008hk}, where no
enhancement was found either. Concerning the
degeneracy of $\eta'$ with its chiral partner f$_0$, we see that such a
degeneracy is achieved only at $T\simeq 300$~MeV. This a consequence of
the strange quark contents of these mesons, which, as we will see below,
become larger as the temperature increases.

\begin{figure}[htb]
\includegraphics[width=0.9\textwidth]{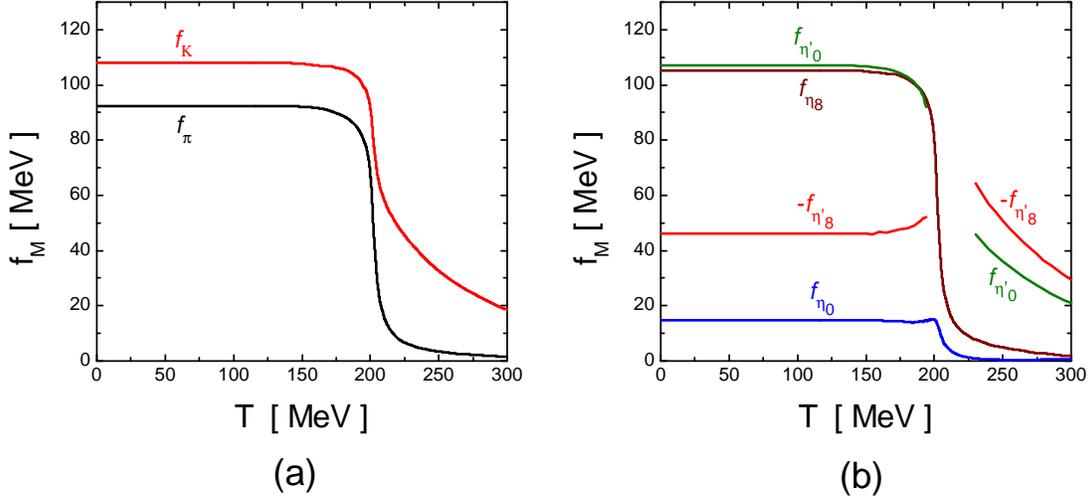}
\caption{Behavior of pseudoscalar meson decay constants as functions of
the temperature.}
\end{figure}
\begin{figure}[htb]
\includegraphics[width=0.9\textwidth]{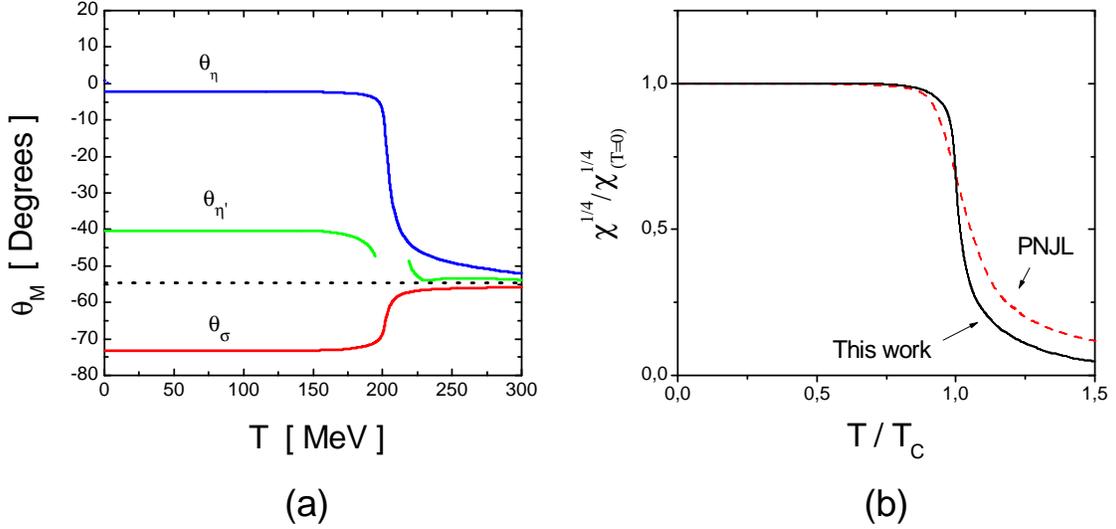}
\caption{(a) Behavior of meson mixing angles as functions of the
temperature. The dashed line shows the ``ideal'' mixing angle $\theta_{\rm
ideal}=\tan^{-1}\sqrt{2}$. (b) Behavior of the topological susceptibility relative
to its $T=0$ value as
function of the temperature in the nonlocal model (solid) and in the
PNJL SU(3) model \cite{Costa:2008dp}~(dotted).}
\end{figure}
\begin{figure}[htb]
\includegraphics[width=0.9\textwidth]{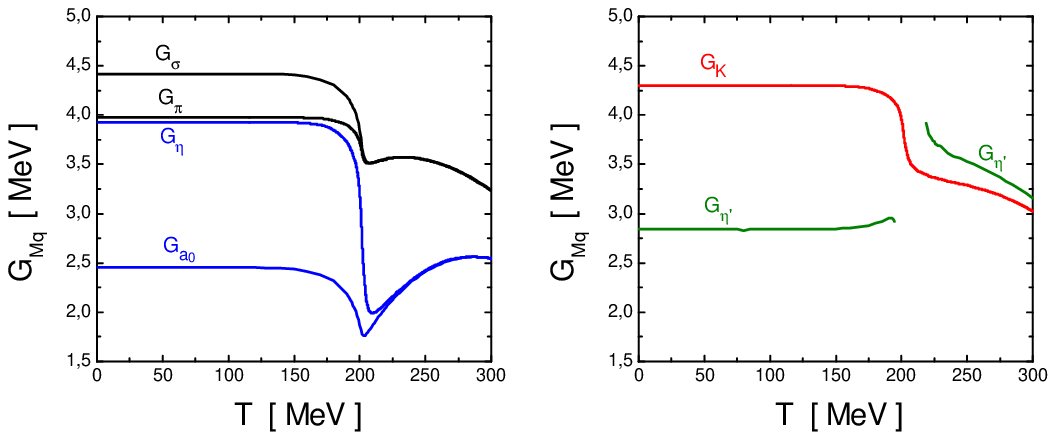}
\caption{Behavior of quark-meson couplings as functions of the
temperature.}
\end{figure}

Next, in Fig.~2 we quote the behavior of pseudoscalar meson decay
constants, which can be calculated from Eqs.~(\ref{fpi}), (\ref{fk}) and
(\ref{faetas}). Fig.~2(a) shows the curves corresponding to the decay
constants $f_\pi$ and $f_K$. It is seen that both decay constants drop
at the phase transition. We observe, however, that due to the strange quark
content of the kaon the corresponding decay constant shows a slower decrease
after the transition.
The behavior of the decay constants associated with $\eta-\eta'$
system is shown in Fig.~2(b). In the case of $f_{\eta_0}$ and
$f_{\eta_8}$ such a behavior is similar to that of $f_\pi$,
while the decrease after the transition is less pronounced
for $f_{\eta'_{0,8}}$. Again, this behaviour of the $\eta'$ decay
constants can be understood in terms of its larger strange quark content.
Here we have left blank the range in which the
$\eta'$ mass is not well defined.

In Fig.~3(a) we plot the behavior of the mixing angles
$\theta_\eta$ and $\theta_{\eta'}$, which can be calculated from
Eq.~(\ref{thetap}). It is seen that above the phase transition
both angles tend to a common value, which is natural since meson
masses also tend to unify. More interestingly, they converge to
the so-called ``ideal'' mixing angle $\theta_{\rm
ideal}=\tan^{-1}\sqrt{2}\simeq 54.7^\circ$ (dashed line in the
figure). This means that, as suggested above, the $\eta$ meson
becomes approximately non-strange, while $\eta'$ approaches to an
$\bar ss$ pair. The same happens with the $\sigma-{\rm f}_0$ pair
(in the figure we have quoted only $\theta_\sigma$, since the
f$_0$ meson mass lies above the $\bar qq$ threshold). The fact
that the mixing angles go to the ``ideal'' value for large
temperatures implies that the U(1)$_A$ anomaly tends to vanish in
this limit. Another signature of this fact is that axial chiral
partners $(\pi, \eta)$ and $(\sigma, a_0)$ become almost
degenerate at $T \simeq 300$~MeV. However, perhaps the best
indication of the vanishing of the U(1)$_A$ anomaly is provided by
the topological susceptibility $\chi$ which, in pure color SU(3)
theory, is related to the $\eta'$, $\eta$ and $K$ masses through
the Witten-Veneziano formula
\begin{equation}
\frac{6}{f_\pi^2} \ \chi \ = \ m_{\eta'}^2 + m_{\eta}^2 - 2 m_K^2
\ .
\end{equation}
Various existing lattice calculations~\cite{Alles:1996nm} show a sharp
decrease of $\chi$ at the critical temperature. In our framework, the
topological susceptibility can be calculated from
\begin{eqnarray}
\!\!\!\!\!\!\!\!\!\! \chi &=& -\frac{H^2}{8}
\left\{ 2\, C^-_{uu} \bar S_u^2 \bar S_s^2\, + C^-_{ss} \bar S_u^4\, -\,
\frac{2}{3} \left[ \frac{\bar S_u (\bar S_u + 2 \bar S_s)}{\sqrt2}
\left(%
\begin{array}{c}
  C^-_{08} \\
  C^-_{00}\\
\end{array}%
\right)^\dagger + \bar S_u(\bar S_s - \bar S_u)
\left(%
\begin{array}{c}
  C^-_{88} \\
  C^-_{80}\\
\end{array}%
\right)^\dagger \right]
\right. \nonumber \\
& & \left. \qquad \qquad \qquad \qquad \qquad
 \cdot\; {\cal G}^{-1}\; \cdot\,
\left[ \frac{\bar S_u (\bar S_u + 2 \bar S_s)}{\sqrt2}
\left(%
\begin{array}{c}
  C^-_{08} \\
  C^-_{00}\\
\end{array}%
\right)
+ \bar S_u (\bar S_s - \bar S_u)
\left(%
\begin{array}{c}
  C^-_{88} \\
  C^-_{80}\\
\end{array}%
\right) \right]\right\} \ ,
\end{eqnarray}
where ${\cal G}$ is a $2\times 2$ matrix whose matrix elements are given in
Eq.~(\ref{g00}), and all functions are evaluated at $(\vec q \ ^2 , \nu_m^2)
= (0,0)$. This expression has been obtained following similar steps as those
described in Ref.~\cite{Fukushima:2001hr} for the case of the (local) NJL
model. The numerical results are shown in Fig.~3(b). To be able to compare
with the result obtained in the local PNJL SU(3) model~\cite{Costa:2008dp}
(where, as already mentioned, $T_c$ turns out to be too high), we show the
normalized value of $\chi^{1/4}$ as a function of $T/T_c$. For both models
one finds a sharp decrease in the topological susceptibility at the critical
temperature, this decrease being steeper in the nonlocal model. Indeed, at
$T/T_c = 1.5$ the ratio $\chi^{1/4}/\chi_{(T=0)}^{1/4}$ is about 11\% for
the local PNJL model, while for the nonlocal model it is roughly one half of
this value. The value of $\chi_{(T=0)}^{1/4}$ is found to be about 162~MeV
in the nonlocal model, while one gets $\simeq 180$~MeV in the PNJL. Recent
lattice calculations (see Ref.~\cite{Durr:2006ky} and references therein)
indicate that $\chi_{(T=0)}^{1/4} \simeq 190$ MeV in pure gauge theories.
However, light dynamical quarks are expected to suppress the topological
susceptibility~\cite{AliKhan:2001ym}. For example, in the lattice
calculation carried out in Ref.~\cite{Alles:2000cg} the authors find
$\chi_{(T=0)}^{1/4} \simeq 163$ MeV for a two-flavor case in the region
where the current quark masses are around 20 MeV.

For completeness, we conclude our description by quoting in Fig.~4 the
behavior of the quark-meson couplings $G_{Mq}$. These can be calculated
from Eqs.~(\ref{zp1}) and (\ref{gpq}) for pseudoscalar mesons, and similar
relations hold for the scalar meson sector with the appropriate changes in
the functions $C_{ij}(\vec q\ ^2,0)$.

\section{Summary and conclusions}

In the present work we have studied the finite temperature behavior of
light scalar and pseudoscalar meson properties in the context of
three-flavor nonlocal chiral models that include mixing with active
strangeness degrees of freedom. The effect of gauge interactions has been
introduced by coupling the quarks with a background gauge field, and the
deconfinement transition has been studied through the behavior of the
traced Polyakov loop. For a given parameterization of the nonlocality
---which, for simplicity, here is introduced through an exponential form
factor---, at zero temperature the model has five free parameters. We have
chosen to fix the average non-strange quark mass $\bar m$ to a
phenomenologically sound value of $\bar m = 5$ MeV, whereas the remaining
four parameters have been determined by requiring that the model
reproduces correctly the measured values of the masses of the pion, kaon
and $\eta'$ pseudoscalar mesons, and the pion decay constant $f_\pi$.
Using this set of parameters one can obtain a very good description of the
remaining zero temperature pseudoscalar meson properties, as well as
adequate values for the scalar meson masses.

In the extension to finite temperature the former parameter values have
been kept fixed, while those appearing in the Polyakov loop potential have
been taken from a fit to lattice results. As expected, the model shows a
fast crossover phase transition, corresponding to the restoration of SU(2)
chiral symmetry. The transition temperature (defined as the position of
the peak of the corresponding chiral susceptibility) is found to be $T_c =
202$~MeV. This value is in better agreement with lattice results,
namely $T_c^{\rm (latt)} = 160-200$~MeV~\cite{Bernard:2004je}, than the
value recently obtained in the local SU(3) PNJL model, $T^{(\rm PNJL)}_c =
259$ MeV~\cite{Costa:2008dp}. In addition one finds a deconfinement phase
transition, which occurs at about the same critical temperature.
Concerning the behavior of meson masses with temperature, it is seen that
pseudoscalar meson masses remain approximately constant up to $T_c$, while
scalar meson masses begin to drop at about 150~MeV. Beyond $T_c$
pseudoscalar masses get increased, in such a way that they become
degenerate with the masses of their chiral partners, as expected from
chiral restoration. The temperature at which chiral partners meet depend
on the strange quark composition of the corresponding mesons, i.e. the
masses of mesons containing no strange quarks match almost immediately
after $T_c$, while f$_0$ and $\eta'$ masses meet only at about 1.5 $T_c$,
the situation being intermediate for K and $\kappa$ mesons. Regarding the
properties of the $\eta-\eta'$ sector, it is seen that the corresponding
mixing angles tend to converge to the so-called ``ideal'' mixing, which
indicates that the U(1)$_A$ anomaly tends to vanish as the temperature
increases. This is also seen in the behavior of the topological
susceptibility which, as expected from lattice calculations, shows a sharp
decrease after the chiral phase transition. It should be noticed, however,
that in the present nonlocal model such a decrease is faster than that
obtained in the local PNJL SU(3)~\cite{Costa:2008dp}. Finally, we notice
that, in agreement with the local model ---and in contrast with what was
suggested in the framework of a Dyson-Schwinger
approach~\cite{Horvatic:2007qs}--- we do not observe any kind of
enhancement of the $\eta'$ mass around the critical temperature.

\section*{Acknowledgements}

This work was supported by CONICET (Argentina) under grants \# PIP 02368
and PIP 02495, and by ANPCyT (Argentina) under grants \# PICT 04-03-25374
and 07-03-00818.

%\pagebreak

\end{document}